\documentclass[a4paper,11pt]{article}
\pdfoutput=1 

\usepackage{jheppub} 

\usepackage[T1]{fontenc} 

\usepackage{pdfsync}
\usepackage{mathptmx}
\usepackage[utf8]{inputenc}
\usepackage[T1]{fontenc}
\usepackage{slashed}
\usepackage{times}
\usepackage{amssymb,amsfonts,amsmath,amsthm}
\usepackage{dsfont,bbm}
\usepackage{dcolumn}
\usepackage{epsf}
\usepackage{graphicx}
\usepackage[caption=false]{subfig}
\usepackage{dsfont}
\usepackage{bm}
\usepackage{tikz}
\usepackage[matrix, arrow, curve]{xy}
\usepackage{slashed}
\usepackage[active]{srcltx}

\title{\boldmath The Gorishny-Isaev vacuum integrations and UV(IR)-regime}

\author[a]{I.~V.~Anikin}

\affiliation[a]{Bogoliubov Laboratory of Theoretical Physics JINR, 141980 Dubna, Russia}

\emailAdd{anikin@theor.jinr.ru}

\abstract{We present the further development of the vacuum massless integrations.
In particular, in the Gorishny-Isaev formula, it has been shown that the delta function 
representing UV-regime should be treated within the sequential approach. 
It allows us to resolve the problem of vacuum integrations related to 
the analytical continuation of diagram indices.

\vspace{4cm}
Based on the materials presented as a talk at 
The XXVth International Baldin-Burov Seminar on High Energy Physics Problems,
September, 21st 2023,
Dubna
}
\begin{document}
\maketitle
\flushbottom

\section{Introduction}

The effective potential computation with multi-loop accuracy 
in the different quantum field models 
plays an very important role for the studies, for example,  
of the spontaneous symmetry breaking manifestation.
As well-known, the calculation of effective potential involves ultimately the 
vacuum integrations. 
The presence of massive parameters (or particle masses) in the theory can significantly
complicate the multi-loop calculations.
On the other hand, the massless propagators in the corresponding loops simplify any
multi-loop calculations.  
Moreover, the massless loops give a possibility to use of conformal symmetry \cite{Anikin:2023wkk,Anikin:2023ogb}.
The price to be paid for the simplification is
the necessity of operation with the $\delta(0)$-singularity after the use of Gorishny-Isaev (GI) vacuum integrations
\cite{Gorishnii:1984te}.  

In the paper, we demonstrate that $\delta(0)$-singularity can be treated as 
either the ultraviolet (IV) or infrared (IR) regime.
To avoid working with the singular generated function, 
we also advocate the use of sequential approach to such kinds of generated functions.
It allows us to resolve a problem with the corresponding analytical extension of 
the GI vacuum integration up to the real indices of diagrams, see \cite{Gorishnii:1984te, Anikin:2020dlh}.

Schematically, the process of effective potential computation can be presented as 
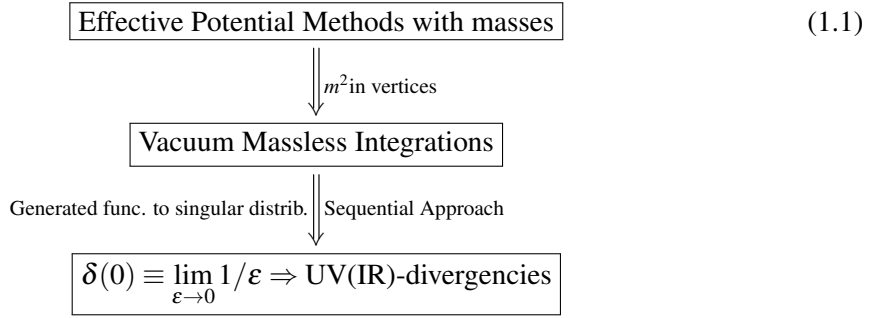
\begin{eqnarray}
\hspace{1cm}
\xymatrix{
&\boxed{\text{Effective Potential Methods with masses}} \ar@{=>}[d]^{m^2 \text{in vertices}}&\\
&\boxed{\text{Vacuum Massless Integrations}}
\ar@{=>}[d]^{\text{Sequential Approach}}_{\text{Generated func. to singular  distrib.}}&\\
&\boxed{\delta(0)\equiv\lim_{\varepsilon\to 0} 1/\varepsilon \Rightarrow \text{UV(IR)-divergencies} }
}
\end{eqnarray}

\section{The massless loop integration in dimensional regularization}

We adhere the dimensional regularization where the loop integration measure is defined for 
$D=d-2\varepsilon$.
Due to the dimensional analysis, the vacuum massless integration gives
\begin{eqnarray}
\int \, \frac{d^D k}{[k^2]^n}=0 \quad \text{for}\quad \forall\,\, n.
\end{eqnarray}
However, if $n=D/2$ the dimensional analysis arguments do not work.
Nevertheless, we can check that the zero is achieved owing to the 
cancellation, each other, of ultraviolet and infrared divergencies \cite{Grozin:2005yg}. 
As a rule, for the effective potential calculations, we concentrate on only 
the ultraviolet (UV) divergences \cite{Anikin:2023wkk,Anikin:2023ogb}. 
Hence, focusing on the only UV-divergency
we deal with the non-zero contribution which is an object of our consideration.
In the case of the IR-divergency, our argumentation is not much different because 
the IR-regime can be obtained from the UV-regime by the corresponding replacement  
$\varepsilon \to -\varepsilon$.

\section{On the singular generated functions}

It is worth to remind that 
every singular functional is being a limit of the regular functional, {\it i.e.}
\begin{eqnarray}
\label{delta-func}
&&\delta(x)=\lim_{\varepsilon\to 0}\delta_\varepsilon(x) \quad \text{\footnotesize in the function space} 
\nonumber\\
&&\Rightarrow \lim_{\varepsilon\to 0} \int d\mu(x) \delta_\varepsilon(x)  \quad \text{\footnotesize in the functional space},
\end{eqnarray}
where the integration measure $d\mu(x)= dx\, \phi(x)$ involves the restricted (finite) function $\phi(x)$.
In Eqn.~(\ref{delta-func}), the function $\delta_\varepsilon(x)$ can be determined by different ways. For example, we have 
\begin{eqnarray}
\label{delta-eps-1}
\delta_\varepsilon(x) = \Big\{   
\frac{1}{\pi} \frac{\varepsilon}{\varepsilon^2 + x^2}; \quad 
\frac{1}{2 \sqrt{\pi\varepsilon}} e^{-\frac{x^2}{4 \varepsilon}};\quad
\frac{1}{\pi} \frac{\sin x/\varepsilon}{x}
\Big\}.
\end{eqnarray}
If the argument of delta function is equal to zero, then we get the following behaviour:
\begin{eqnarray}
\label{delta-eps-2}
\delta_\varepsilon(x)
\stackrel{x=0}{\thicksim}
\Big\{
\big[\frac{1}{\varepsilon}]; \quad
\big[ \frac{1}{\sqrt{\varepsilon}}];\quad
\big[ \frac{1}{\varepsilon}]
\Big\},
\end{eqnarray}
where $\stackrel{x=0}{\thicksim}$ means ``behaves as'' if $x=0$.
One can see that Eqn.~(\ref{delta-eps-2}) gives the different parametrizations of the $\delta(0)$-singularity.  
Actually, it is well-known that the $\delta(0)$-singularity can be presented through the different meromorphic 
(and not only) functions \cite{Gelfand:1964}.

\section{ $\Delta_F(0)$-singularity}

As the first example, let us consider the tad-pole diagrams which, by definition, include 
the singularity as $\Delta_F(0)$. 
Using the Fourier transform, the propagator $\Delta_F(0)$ can be write as
\begin{eqnarray}
\label{V-Int-1}
\Delta_F(0) = \int \frac{(d^D k)}{k^2}=
\Gamma(D/2-1) \int (d^Dz) \, \,\frac{\delta(z)}{\big(z^2\big)^{D/2-1}},
\end{eqnarray}
where the causal prescription $+i0$ is omitted in the denominators.
If we assume that $D/2-1=0$, then 
\begin{eqnarray}
\label{V-Int-2}
\Delta_F(0) = \Gamma(0) \int (d^Dz) \, \delta(z) \Rightarrow \Gamma(0),
\end{eqnarray}
where
\begin{eqnarray}
\label{V-Int-3}
 \Gamma(0) = \lim_{\epsilon\to 0} \Gamma(\epsilon)=  \lim_{\epsilon\to 0} \Big\{\frac{1}{\epsilon} + ....\Big\}.
\end{eqnarray}
The condition given by $D/2-1=0$ has to
be applied before the integration over $(d^D k)$ in order to avoid the uncertainty.

On the other hand, the vacuum integration 
results in the delta function \cite{Gorishnii:1984te}. 
A key moment of Gorishny-Isaev's method goes as below.
Using the spherical system, $\Delta_F(0)$ can be represented as
\begin{eqnarray}
\label{V-Int-1-w1}
&&\Delta_F(0) = \int \frac{(d^D k)}{k^2}=
\frac{1}{2} \int d\Omega \int_{0}^\infty d\beta \, \beta^{D/2-2}.
\end{eqnarray}
The replacement $\beta = e^y$ leads to the following expression
\begin{eqnarray}
\label{V-Int-1-w2}
\Delta_F(0) =
\frac{1}{2} \int d\Omega \int_{-\infty}^\infty (dy) \, e^{iy \big[(-i)(D/2-1)\big]}
\stackrel{n}{=}
\frac{1}{2}\delta\big( i[D/2 -1] \big)=
\frac{1}{2\, | i |}\delta\big( D/2 -1 \big).
\end{eqnarray}
The angular integration with $d\Omega$ gives only the finite numerical coefficient.

It is worth to notice that the representation given by Eqn.~(\ref{V-Int-1-w2}) is valid if and only if 
the delta function argument is a complex number.

Now, restoring all coefficients, it reads
\begin{eqnarray}
\label{V-Int-3-2}
\Delta_F(0) =- \, 2i\, \pi^{1+D/2} \,\delta(1-D/2) \Big|_{D=2} = - \, 2i\, \pi^{2} \,\delta(0).
\end{eqnarray}
So, for the case of $D=2$,  owing to the corresponding matching, one can observe that the singularities given by 
two different functions $\Gamma(0)$ and $\delta(0)$ can be parametrized in the same way, {\it i.e.}
\begin{eqnarray}
\label{V-Int-4}
(-i) \,  \Delta_F(0)=\Gamma(0) = - \, 2\, \pi^{2} \,\delta(0).
\end{eqnarray}
With this, we may conclude that $\delta(0)$-singularity can be treated as the singularity of $\Gamma(0)$.
The same inference has been reached by the different method, see \cite{Anikin:2020dlh}.
Notice that the physical nature of the mentioned singularity has been somewhat hidden.

\section{$\delta(0)$-singularity and the sequential approach}

As above-mentioned, based on the dimensional analysis, we may conclude that 
\begin{eqnarray}
\label{V-in-1}
\mathcal{V}_n=\int \frac{(d^D k)}{[k^2]^n}=0 \quad \text{for}\,\, n\not= D/2.
\end{eqnarray}
The case of $n=D/2$ (or $n=2$ if $\varepsilon\to 0$) requires the special consideration because
the dimensional analysis argumentation does not now work.
The nullification of $\mathcal{V}_{D/2}$ takes still place but thanks to different reasons.
It turns out, UV- and IR- divergencies are cancelled each other.
Hence, if only the ultraviolet divergencies are under our consideration, $\mathcal{V}_{D/2}$ is not equal to zero.

To demonstrate it, we dwell on the vacuum integration where 
the IR-regularization has been applied. Notice that the mentioned IR-regularization is 
needed as the external operation to be focused on the UV-regime only.
In the
spherical co-ordinate system, we write the following 
representation 
\footnote{$\mu^2$ plays a role of IR-regularization.}
\begin{eqnarray}
\label{V-in-2}
\mathcal{V}_{2}=\int_{UV} \frac{(d^D k)}{[k^2]^2}\equiv
 \frac{\pi^{D/2}}{\Gamma(D/2)} \int_{\mu^2}^{\infty} d\beta \beta^{D/2-3}
\,\,\,\, \text{with}\,\,\, \beta=|k|^2,
\end{eqnarray}
Next, calculating $\beta$-integration, we reach the representation as
\begin{eqnarray}
\label{V-in-3}
\mathcal{V}_{2}=
 \frac{\pi^{2-\varepsilon} \mu^{-2\varepsilon} }{\Gamma(2-\varepsilon)}  \, \frac{1}{\varepsilon} \Big|_{\varepsilon\to 0},
\end{eqnarray}
where the $\epsilon$-pole corresponds to the UV-divergency only (the IR-divergency is absent
by construction thanks for $\mu^2$) \cite{Grozin:2005yg}.

On the other hand, we are able to calculate the vacuum integration by
Gorishny-Isaev's method \cite{Gorishnii:1984te}. In this case, $\mathcal{V}_n$ reads
\begin{eqnarray}
\label{V-in-4}
\mathcal{V}_{n}=\int \frac{(d^D k)}{[k^2]^n} =
\frac{2i\, \pi^{1+D/2}}{(-1)^{D/2}\, \Gamma(D/2)} \delta(n-D/2).
\end{eqnarray}
Supposing $D=4-2\varepsilon$, the only contribution is given by
\begin{eqnarray}
\label{V-in-4-2}
\mathcal{V}_{2}=\int \frac{(d^D k)}{[k^2]^2} =
\frac{2i\, \pi^{3-\varepsilon}}{ \Gamma(2-\varepsilon)} \delta(\varepsilon) \, \not=\, 0.
\end{eqnarray}
Hence, the delta function of argument $\epsilon$ reflects, in fact, the UV-divergency
because the representations of $\mathcal{V}_2$ given by Eqns.~(\ref{V-in-3}) and (\ref{V-in-4-2})
should be equivalent.

In terms of generated functions (functionals), the delta function is a linear singular 
functional  defined on the space of finite basic functions.
We remind that a singular generated function cannot be generated by the locally-integrated functions.
Also, the delta function can be understood with the help of the fundamental 
sequences of regular functionals provided the corresponding weak limit, see Eqn.~(\ref{delta-func}) \cite{Gelfand:1964, Antosik:1973}.
Besides, one of the delta function representations is related to the following realization
\begin{eqnarray}
\label{Delta-Real}
\delta(t)=\lim_{\varepsilon\to 0} \delta_\varepsilon(t)\equiv
\lim_{\varepsilon\to 0}  \frac{ \Theta(-\varepsilon \le t \le 0)}{\varepsilon},
\end{eqnarray}
where $\Theta(-\varepsilon \le t \le 0)$ implies the extended step-function without any uncertainties.
This representation of delta function augments the set of possible representations, see Eqn.~(\ref{delta-eps-1}).

One can see that the treatment of $\delta(\varepsilon)$ as the linear (singular)
functional on the finite basic function space with $d\mu(\varepsilon)=d\varepsilon \phi(\varepsilon)$ meets some difficulties
within the dimensional regularization approach. Indeed, for the practical use,
$\varepsilon$ is not a convenient variable for the construction of the finite function space because we finally need
to be focused on the limit as $\epsilon\to 0$.

Meanwhile, within the sequential approach \cite{Antosik:1973, Gelfand:1964},
the delta function might be considered as the usual singular (meromorphic)
function and  the $\delta(0)$-singularity can be treated as
a pole of the first order \cite{Anikin:2020dlh},
\begin{eqnarray}
\label{D-treat}
\delta(0)=\lim_{\varepsilon\to 0}  \delta_{\varepsilon}(0)\equiv \lim_{\varepsilon\to 0}  \frac{1}{\varepsilon}.
\end{eqnarray}
For the demanding mathematician, this representation should be understood merely as a symbol.
That is, $\delta(0)$ denotes alternatively the limit of $1/\epsilon$.

Generally speaking, the parametrization of $\delta(0)$-singularity can be realized by many different ways,
see Eqns.~(\ref{delta-eps-2}) and (\ref{D-treat}).
However, in our consideration, the unique parametrization in the form of Eqn.~(\ref{D-treat})
can be fixed by the fact that Eqns.~(\ref{V-in-3}) and (\ref{V-in-4-2}) are equivalent ones.

Thus, in our study, the $\delta(0)$-singularity is parametrized in the form of the meromorphic function (see Eqn.~(\ref{D-treat}))
within the sequential approach \cite{Antosik:1973, Gelfand:1964}.
Compared to Eqn.~(\ref{V-Int-1-w2}), the representation of Eqn.~(\ref{D-treat}) together with Eqn.~(\ref{V-in-3})
are not limited by the complex argument of 
delta function only. This is an obvious preponderance of the sequential approach. 

\section{Conclusions}

To conclude, we have presented the important extension  of the Gorishny-Isaev massless vacuum integrations.
In the note, we have demonstrated the preponderance of sequential approach where
the singular generated functions (distributions) are treated as a fundamental
sequences of regular functionals. 
Due to this treatment, the singularity as $\delta(0)$ can be parametrized with the help of
the meromorphic function of first order.
The uniqueness of parametrization is shown to be ensured by the fact that 
the delta function represents the UV-regime.

\section*{Acknowledgements}

We thank M.~Hnatic, A.~Manashov, S.V.~Mikhailov and L.~Szymanowski for very useful discussions.


\begin{thebibliography}{99}
\vspace{1\baselineskip}


\bibitem{Anikin:2023wkk}
I.~V.~Anikin,
``Conformal Symmetry and Effective Potential: I. Vacuum $V_{z,x}$-operation for the Green functions,''
[arXiv:2306.15373 [hep-ph]].

\bibitem{Anikin:2023ogb}
I.~V.~Anikin,
``Conformal Symmetry and Effective Potential: II. Evolution,''
[arXiv:2306.17018 [hep-ph]].

\bibitem{Gorishnii:1984te}
  S.~G.~Gorishnii and A.~P.~Isaev,
  Theor.\ Math.\ Phys.\  {\bf 62}, 232 (1985)
  [Teor.\ Mat.\ Fiz.\  {\bf 62}, 345 (1985)].

\bibitem{Anikin:2020dlh}
I.~V.~Anikin,
Phys.\ Part.\ Nucl.\ Lett.\ {\bf 18}, 290 (2021)

\bibitem{Grozin:2005yg}
A.~Grozin,
``Lectures on QED and QCD,''
[arXiv:hep-ph/0508242 [hep-ph]].


\bibitem{Gelfand:1964}
 I.~M.~Gelfand and G.~E.~Shilov,
``Generalized Functions Vol 1 Properties And Operations,''
Academic Press, 1964, ISBN-0-12-279501-6


\bibitem{Antosik:1973}
Antosik P., Mikusinsky Y., Sikorsky R., ``Theory of Generalized Functions: A Sequential Approach,''
(PWN -- Polish Scientific Publisher, 1973)



\end{thebibliography}
\end{document}